\documentclass[aps,amssymb,floatfix,prd,preprintnumbers]{revtex4-2}
\usepackage{epstopdf}
\usepackage{capt-of}
\usepackage{graphicx}  
\usepackage{dcolumn}   
\usepackage{bm}
\usepackage{amsmath}
\usepackage[font=scriptsize]{caption}
\usepackage[colorlinks]{hyperref}
\begin{document}
\input epsf.tex
\title{\bf Gravitational baryogenesis models comparison in $f(R)$ Gravity}

\author{A.S. Agrawal \footnote{Department of Mathematics, Birla Institute of Technology and Science-Pilani, Hyderabad Campus, Hyderabad-500078, India, email:agrawalamar61@gmail.com}, S.K. Tripathy \footnote{Department of Physics, Indira Gandhi Institute of Technology, Sarang, Dhenkanal, Odisha-759146, India, tripathy\_sunil@rediffmail.com}, B.Mishra \footnote{Department of Mathematics,
Birla Institute of Technology and Science-Pilani, Hyderabad Campus,
Hyderabad-500078, India, email: bivu@hyderabad.bits-pilani.ac.in}}

\affiliation{ }

\begin{abstract}
\textbf{Abstract}
We have studied the gravitational baryogenesis in $f(R)$ theory of gravity with an anisotropic Bianchi I space-time. The matter field is considered to be that of perfect fluid. Two models pertaining to specific form of Ricci scalar have been presented. The baryon-to-entropy ratio has been derived with the some specific form of Ricci scalar in the anisotropic background. The gravitational baryogenesis is examined and its behaviors are studied. 
\end{abstract}

\maketitle

\section{Introduction}
The absence of intense radiation from matter-anti matter annihilation \cite{Cohen98}, the findings on the big-bang nucleosynthesis (BBN) \cite{Burles01} and the Cosmic Microwave Background (CMB) \cite{Bennet03} have indicated that the Universe contains excess matter than anti-matter. The reason on the generation of baryon asymmetric still remains as an open problem. However, the finding from the measurements of CMB anisotropy on small angular scales prescribes the baryon density accuracy of around 1\%. According to the Sakharov criteria \cite{Sakharov67, Sakharov1991, Sakharov1991a}, three conditions are necessary to generate the baryon asymmetry such as (i) baryon-number non-conserving interactions; (ii) ${\mathcal{C}}$ and ${\mathcal{C}}P$ violation and (iii) the out of thermal equilibrium. It has been revealed that \cite{Sakharov67} the Universe was contracting initially with equal and opposite asymmetry. Then it bounced at the singularity and reversed the magnitude of its baryon symmetry. It can be noted that baryogenesis (BG) is a mechanism to generate the baryon asymmetry in the Universe. This depends on the coupling between the baryon matter current $J^{\mu}$ and the Ricci scalar curvature $R$. The spontaneous baryogenesis (SBG) has been proposed with the features of generation of baryon asymmetry in thermal equilibrium without the necessity of  ${\mathcal{C}}$ or ${\mathcal{C}}P$ violation \cite{Cohen1987, Cohen1991}. The gravitational baryogenesis (GBG) has been proposed by Davoudiasl et al. to explain the excess of matter over anti-matter in the observable Universe \cite{Davoudiasl2004}. Davoudiasl et al. \cite{Davoudiasl2004} have shown that the gravitational interaction between the baryon-number current and the derivative of Ricci scalar curvature dynamically breaks the $\mathcal{CPT}$ symmetry in an expanding Universe. Since the proposal of Davoudiasl et al., the gravitational baryogenesis has gained a lot of popularity and many generalizations of the original mechanism have been proposed \cite{Lambiase2006, Lambiase2013, Odintsov2016, Fukushima2016, Simone2017, Arbuzova2017}.

The gravitational baryogenesis for baryon and antibaryon asymmetry is obtained on the basis of the presence of a $\mathcal{CP}$-violating interaction term of the form \cite{Davoudiasl2004}
\begin{equation} \label{eqn.1}
\frac{\epsilon}{M_{*}^{2}}\int d^{4}x\sqrt{-g}(\partial_{\mu}R)J^{\mu}
\end{equation}
where, $\epsilon=\pm1$, $g$ be the metric determinant and $M_{*}$ is the parameter that represents the cut-off scale that characterizes the effective gravitational theory. It should be mentioned here that, this term in \eqref{eqn.1} may occur from higher order interactions in gravitational physics controlling the high energy regime. The baryon number density $n_b$  in thermal equilibrium may be expressed as \cite{Davoudiasl2004} 

\begin{equation}\label{eqn.2}
n_b=n_B-n_{\bar{B}}=\frac{g_bT^3}{6\pi^2}\left(\frac{\pi^2 \mu_B}{T}+\left(\frac{\mu_B}{T}\right)^3\right),
\end{equation} 
where $g_{b}$ denotes the number of intrinsic degree of freedom of baryons, $g_{b}\simeq 1$, $\mu_B$ is the chemical potential and $\mu_B=-\mu_{\bar{B}}=-\epsilon \dot{R}/M_*^2$ and $T$ is the temperature. The baryon number to entropy ratio can be defined as
\begin{equation} \label{eqn.3}
\frac{n_{b}}{s}\simeq -\epsilon \frac{15g_{b}}{4\pi ^{2}g_{s}}\frac{\dot{R}}{M_{*}^{2}T}\bigg\vert_{T_{D}}
\end{equation} 
where $g_{s}\simeq 106$ denotes the number of degree of freedom of effective massless particle and $s=(2\pi^{2}/45)g_{s}T^{3}$. $T_D$ represents the temperature in which the baryon current decouples. Whenever the universe expands and cools, the temperature drops below $T_D$ and the asymmetry can no longer change and is frozen. Now, the net asymmetry remains to be, $\frac{n_{b}}{s}\simeq \frac{\dot{R}}{M_{*}^{2}T}\bigg\vert_{T_{D}}$. We have assumed the cut-off scale to take the value, $M_{*}=10^{12}$ GeV and the critical temperature, $T_D=2\times 10^{16} GeV$ \cite{Lambiase2006}. The observational constraint on the value of baryon-to-entropy ratio is $\frac{n_b}{s}=9.2^{+0.6}_{-0.4}\times 10^{-11}$ \cite{Bennet03}.  For a flat FRW Universe, the baryon-to-entropy ratio is proportional to $\dot{R}$ and it becomes zero for a radiation dominated relativistic matter fluid with the equation of state parameter $\omega=1/3$. \\  

In the last two decades, several observations approved the claim of accelerated expansion of the universe at least at the late phase of cosmic evolution. \cite{Riess98,Perlmutter99,Spergel03,Ade16}. Theoretically, these findings could not be well answered by Einstein's General Relativity (GR). Therefore, the need of a modification in GR has become inevitable. The modification can be in terms of an extension of GR or to formulate new gravity theory. However, any new gravity proposed in recent past is either a modification or an extension of GR. $f(R)$  gravity is such an extended gravity theory which has some compatibility to study this issue \cite{Carroll04}.  The action for $f(R)$ gravity is written as

\begin{equation} \label{eqn.4}
S=\frac{1}{2\kappa^2}\int d^4x\sqrt{-g}f(R)+\int d^4x\sqrt{-g}\mathcal{L}_m
\end{equation}
where, $\kappa^2=\frac{8\pi G}{c^4}$, $\mathcal{L}_m$ is the matter Lagrangian density.  The field equations of the $f(R)$ gravity can be obtained from the action \eqref{eqn.4} as,

\begin{equation} \label{eqn.5}
\frac{\partial f(R)}{\partial R} R_{\mu \nu}-\frac{1}{2}f(R)g_{\mu \nu}-\nabla_{\mu}\nabla_{\nu}\frac{\partial f(R)}{\partial R}+g_{\mu \nu}\square \frac{\partial f(R)}{\partial R}=kT_{\mu \nu}.
\end{equation}
Here $\square\equiv \nabla_{\mu}\nabla^\mu$ is the d'Alembert operator and $\nabla_{\mu}$ represents the covariant differentiation. We have considered the natural system of unit,  $8\pi G=\hbar=c=1$, $G$ is the Newtonian gravitational constant, $\hbar$ and $c$ respectively represent the reduced Planck constant and speed of light in vacuum respectively. The GR field equations can be recovered from  \eqref{eqn.5} for $f(R)=R$. The energy momentum tensor can be expressed as,

\begin{equation}\label{eqn.6}
T_{\mu \nu}=-\frac{2}{\sqrt{-g}}\frac{\delta(\sqrt{-g} \mathcal{L}_m)}{\delta g^{\mu\nu}}.
\end{equation}
Contracting \eqref{eqn.5} we obtain  

\begin{equation} \label{eqn.7}
\frac{\partial f(R)}{\partial R}R-2f(R)+3\square \frac{\partial f(R)}{\partial R}=T
\end{equation} 
where $R=g^{ij}R_{ij}$ and $T=g^{ij}T_{ij}$ are respectively the Ricci scalar and trace of stress energy tensor. Eqn. \eqref{eqn.7}, shows that the Ricci scalar $R$ is a fully dynamical degree of freedom \cite{Lobo09}. Sotiriou and Faraoni \cite{Sotiriou10} have presented the formalism of $f(R)$ gravity in metric, Palatini, and metric-affine approaches. They have given a comprehensive study on the action, field equations and the theoretical aspects of these approaches.  
$f(R)$ gravity has been successful in many aspects. Tripathy and Mishra \cite{Tripathy16} have obtained anisotropic solutions in $f(R)$ gravity without invoking any matter field. Mongwane \cite{Mongwane17} has formulated the characteristic initial value problem for $f(R)$ gravity. Liu et al. \cite{Liu18} have constrained the $f(R)$ gravity in cosmology, solar system and pulsar systems. Lazkoz \cite{Lazkoz18} reformulated the Lagrangian terms of $f(R)$ gravity as an explicit functions of the redshift. Many researchers have worked on $f(R)$ gravity in recent past on different aspects of cosmology \cite{Yang09,  Capozziello12,Harko13,Bamba14,Pavlovic15,Odintsov15,Bahamonde16,Nunes17,Eiroa18,Manzoor19,Yousaf20}.  Since $f(R)$ gravity answered some of the key issues of late time cosmic expansion, we are motivated to examine whether this gravity can provide a reason to the occurrence of gravitational baryogenesis that leads to the observed baryon asymmetry in the Universe in an anisotropy background.\\

Several cosmological models are proposed in recent past on the issues of baryogenesis in extended gravity theory. Lambiase and Scarpetta \cite{Lambiase2006} have reviewed the $f(R)$ theory of gravity in gravitational baryogenesis. Oikonomou and Saridakis \cite{Oikonomou16} have investigated the occurrence of  baryogenesis in the gravitational term of $f(T)$ gravity. Ramos and Paramos \cite{Ramos17}, have generalized the mechanism for gravitational baryogensis in the $f(R)$ gravity theory. Nozari and Rajabi \cite{Nojari18} explored the issue of baryon asymmetric with $f(R,T)$ gravity. Aghamohammadi et al. \cite{Aghamohammadi18} have studied the effect of $f(R)$ gravity on the baryon-to-entropy ratio.  Arbuzova and Dolgov \cite{Arbuzova19} have shown the instability effects both in gravitational baryogenesis and in $f(R)$ gravity. Oikonomou et al. have studied the gravitational baryogenesis mechanism for generating baryon asymmetry in the context of running vacuum models and demonstrated that a nonzero baryon-to-entropy ratio may be produced for a Universe filled with conformal matter \cite{Oikonomou2017}. Within the framework of Dvali–Gabadadze–Porrati (DGP) brane cosmology, Atazadeh has investigated the GBG and found that, the  $\frac{n_b}{s}$ ratio is non zero for a radiation dominated Universe \cite{Atazadeh2018}. Antunes et al. have obtained a non-zero $\frac{n_b}{s}$ ratio in the radiation dominated early Universe without considering the $\mathcal{CPT}$ violation \cite{Antunes2019}. The paper is organised as follows. In Sec II, we have derived the field equations of $f(R)$ gravity in Bianchi I space-time and matter field in the form of perfect fluid. The gravitational baryogenesis has been discussed for the specific Ricci scalar in Section III. The results and discussions are presented in section IV.

\section{Field Equations}

In order to examine the occurrence of baryon asymmetry and study the effect of cosmic anisotropy on the baryon number-to-entropy ratio, in the present work, we consider a spatially homogeneous and anisotropic  Bianchi type I (BI) metric as,
\begin{equation} \label{eqn.8}
ds^2=-dt^2+A^2dx^2+B^2(dy^2+dz^2).
\end{equation}
The metric potentials $A$ and $B$ are considered to be functions of cosmic time only. We assume that, $A(t)\neq B(t)$. The BI metric reduces to the usual flat FRW metric for $A(t)=B(t)$. The given BI metric provides us an opportunity to investigate the possible effect of cosmic anisotropy on the dynamical aspects of the model while keeping room to be reduced to the usual FRW Universe. In order to simplify the whole problem and to avoid any issues regarding the occurrence of ghost fields, we consider a perfect fluid distribution in the Universe and left all the burden of the late time cosmic acceleration to the geometry modification through the $f(R)$ gravity.  For a perfect fluid distribution, the energy momentum tensor is expressed as,

\begin{equation} \label{eqn.9}
T_{\mu \nu}=(\rho+p)u_{\mu}u_{\nu}+pg_{\mu \nu},
\end{equation}
where $p$ and $\rho$ respectively denote the pressure and energy density of the matter field. $u^{\mu}$ is the four velocity vector of the fluid satisfying $u^{\mu}u_{\mu}=-1$.   

The space-time \eqref{eqn.8} is plane symmetric and the directional scale factors are different. Also, the rate of expansion along the spatial directions are not equal. Here, we can define the directional Hubble rate for the directional scale factors in \eqref{eqn.8} as, $H_x=\frac{\dot{A}}{A}, H_y=H_z=\frac{\dot{B}}{B}$, where the over dot represents derivative with respect to cosmic time. The mean Hubble parameter may be expressed in terms of the directional Hubble parameters as, $H=\frac{H_x+2H_y}{3}$. The Ricci scalar $R$ can be expressed with respect to the directional Hubble parameters as, 

\begin{equation} \label{eqn.10}
R=-2\left[\dot{H}_x+H_x^2+2\dot{H}_y+3H_y^2+2H_xH_y\right]
\end{equation}
Since shear scalar, $\sigma^2=\sigma_{ij}\sigma^{ij}$ can also be expressed with respect to the metric potential as well to the directional Hubble rates, we express it as, $\sigma^2=\frac{1}{2}(H_x^2+2H_y^2)-\frac{3}{2}H^2$ for the assumed space-time. Now, The field equations of $f(R)$ gravity for the space-time \eqref{eqn.8} and the energy momentum tensor \eqref{eqn.9} in the form of Hubble parameter can be obtained as,

\begin{eqnarray} \label{eqn.11}
\frac{f(R)}{2}-\left(\dot{H}+3H^2\right)f_R+\left(\ddot{R}+2\dot{R}H\right)f_{RR}+\dot{R}^2f_{RRR}=-p\\
\frac{f(R)}{2}+\left[3\left(\dot{H}+3H^2\right)-R\right]f_R+3H\dot{R}f_{RR}=\rho\label{eqn.12}
\end{eqnarray}
The subscript $R$ refers to partial differentiation with respect to the Ricci scalar $R$. In order to solve the above set of field equations \eqref{eqn.11}-\eqref{eqn.12}, either we require a presumed equation of state between the pressure and the energy density or we take some arbitrary functional form of $f(R)$ satisfying the stability conditions. Also in order to maintain some amount of anisotropy among the field variables, we have considered $H_x=\eta H_y$ so that the isotropic model is recovered for  $\eta=1$. In a recent work \cite{Tripathy16}, Tripathy and Mishra have considered some anisotropic solutions in $f(R)$ gravity and have proposed some functional forms for $f(R)$. These functional forms are stable against the conditions $f_R >0$ and $f_{RR}>0$, where $f_{RR}=\frac{d^2f(R)}{dR^2}$. In the present investigation, we wish to consider the functional $f(R)$ in the form $f(R)=\left(\frac{2R}{\gamma+2}\right)\left(\frac{R}{\chi}\right)^{\frac{\gamma}{2}}$ \cite{Tripathy16}, where $\chi=18\left(\frac{1}{\gamma+1}-\frac{\eta^2+2\eta+3}{\eta^2+4\eta+4}\right)$ and $\gamma$ is a constant model parameter. 

\section{Specific $f(R)$ gravity models and gravitational baryogenesis}
In this section, we wish to study the gravitational baryogenesis for the geometrically modified $f(R)$ gravity. Since we have considered an anisotropic Universe with different directional Hubble rates along different directions, we wish to study the effect of cosmic anisotropy on the baryon number-to-entropy ratio $\frac{n_b}{s}$. The expression for $\frac{n_b}{s}$ is given in eqn. \eqref{eqn.3} which depends on the time derivative of the Ricci scalar and the baryon current decoupling temperature $T_D$. In the following discussions, we will consider two different approaches to obtain the baryon current decoupling temperature and then study the influence of the cosmic anisotropy. 
\subsection{Model I}
As a first example, we consider $$R=\chi H^2=\chi\left(\frac{\mathcal{H}_0}{n(t-t_0)+1}\right)^2,$$ such that
\begin{equation}
\dot{R}=-2\chi \frac{n\mathcal{H}_0^2}{(n(t-t_0)+1)^3},\label{eqn.13}
\end{equation}
 where $n$ and $\mathcal{H}_0$ are constants. Now, substituting \eqref{eqn.13} in eqn. \eqref{eqn.3}, we obtain
 \begin{equation}
 \frac{n_b}{s}\simeq\chi\epsilon \frac{15g_b}{2\pi^2g_s}\frac{n\mathcal{H}_0^2}{\left[n(t_D-t_0)+1\right]^3 M_*^2T_D}. \label{eqn.14}
\end{equation}
where $t_D$ denotes the decoupling time. Adiabatic decaying of the vacuum  may be assumed here which ensures a constant specific entropy of the massless particles. Basing on the covariant non-equilibrium thermodynamic description, Lima et al. \cite{Lima96,Lima97} have derived the standard expressions for the energy density $\rho_R$ and number density $n_R$ as the function of the temperature i.e., $\rho_R \varpropto T^4_R$ and $n_R \varpropto T^3_R$. Moreover, the expression for $\rho$ has been suggested as \cite{Aghamohammadi18},
\begin{equation}
\rho=\left(\frac{g_s\pi^2}{30}\right) T^4. \label{eqn.15}
\end{equation}
Substituting this assumption in eqn. \eqref{eqn.12}, we obtain

\begin{equation}
T=\left(\frac{30}{g_{s}\pi^{2}}\right)^{\frac{1}{4}} \left(\frac{R}{\chi}\right)^{\frac{\gamma}{8}}\left[3(\dot{H}+3H^2)-R\left(\frac{\gamma+1}{\gamma+2}\right)-\frac{3 H n \mathcal{H}_{0}^{2} \gamma \chi }{(n(t-t_{0})+1)^{3}R}\right]^{\frac{1}{4}} \label{eqn.16}
\end{equation}
Now using eqn. \eqref{eqn.16} in eqn. \eqref{eqn.14}, we obtain the baryon number-to-entropy ratio as,
\begin{equation}
\frac{n_{b}}{s}\simeq \epsilon\frac{15g_{b}}{2\pi^{2}g_{s}}\frac{n\chi}{M_{*}^{2}}\left(\frac{g_s\pi^2}{30}\right)^{\frac{3}{2+\gamma}}\frac{T_D^{\frac{12}{2+\gamma}-1}}{\mathcal{H}_{0}^{\frac{3\gamma}{\gamma+2}-2}}\left[\frac{\chi \mathcal{H}_{0}^{2}}{\gamma+2}+9\mathcal{H}_{0}^{2}-3n\mathcal{H}_{0}-3\mathcal{H}_{0}n\gamma \right]^{\frac{-3}{2+\gamma}} \label{eqn.17}
\end{equation}

For $ \epsilon= 1$, $g_b \simeq 1$ and $g_s \simeq 106$, eqn. \eqref{eqn.17} reduces to,

\begin{equation}\label{eqn.18}
\frac{n_{b}}{s}\simeq (0.007168)\frac{n\chi }{M_{*}^{2}}(34.83)^{\frac{3}{2+\gamma}}\frac{T_D^{\frac{12}{2+\gamma}-1}}{\mathcal{H}_{0}^{\frac{3\gamma}{\gamma+2}-2}}\left[\frac{\chi \mathcal{H}_{0}^{2}}{\gamma+2}+9\mathcal{H}_{0}^{2}-3n\mathcal{H}_{0}-3\mathcal{H}_{0}n\gamma \right]^{\frac{-3}{2+\gamma}} 
\end{equation}
The graphical representation of the baryon number to entropy ratio has been presented with respect to $n$ and the anisotropic parameter $\eta$ in Fig. 1 and Fig. 2 respectively. Fig. 1 depicts the behavior of $\frac{n_b}{s}$ with respect to $n$ for three representative values of the anisotropic parameter, $\eta=0.95,1.5,2.1$. It can be observed from the figure that, for a given value of the anisotropic parameter $\eta$, $\frac{n_b}{s}$ increases almost linearly from an initial null value. With an increase in the value of the anisotropic parameter, the slope of the $\frac{n_b}{s}$ curve increases slightly for larger values of $n$. In Fig. 2, we have shown the effect of anisotropy on the baryon number-to-entropy ratio for three representative values of $n$ namely $n=0.9,1,1.1$. For a given value of $n$, the $\frac{n_b}{s}$  ratio in general increases with an increase in the anisotropic parameter. It starts from a small positive value and with the increase in the anisotropic parametric value, it further increases in the positive domain. 
\begin{figure}[h!]
\minipage{0.50\textwidth}
\includegraphics[width=75mm]{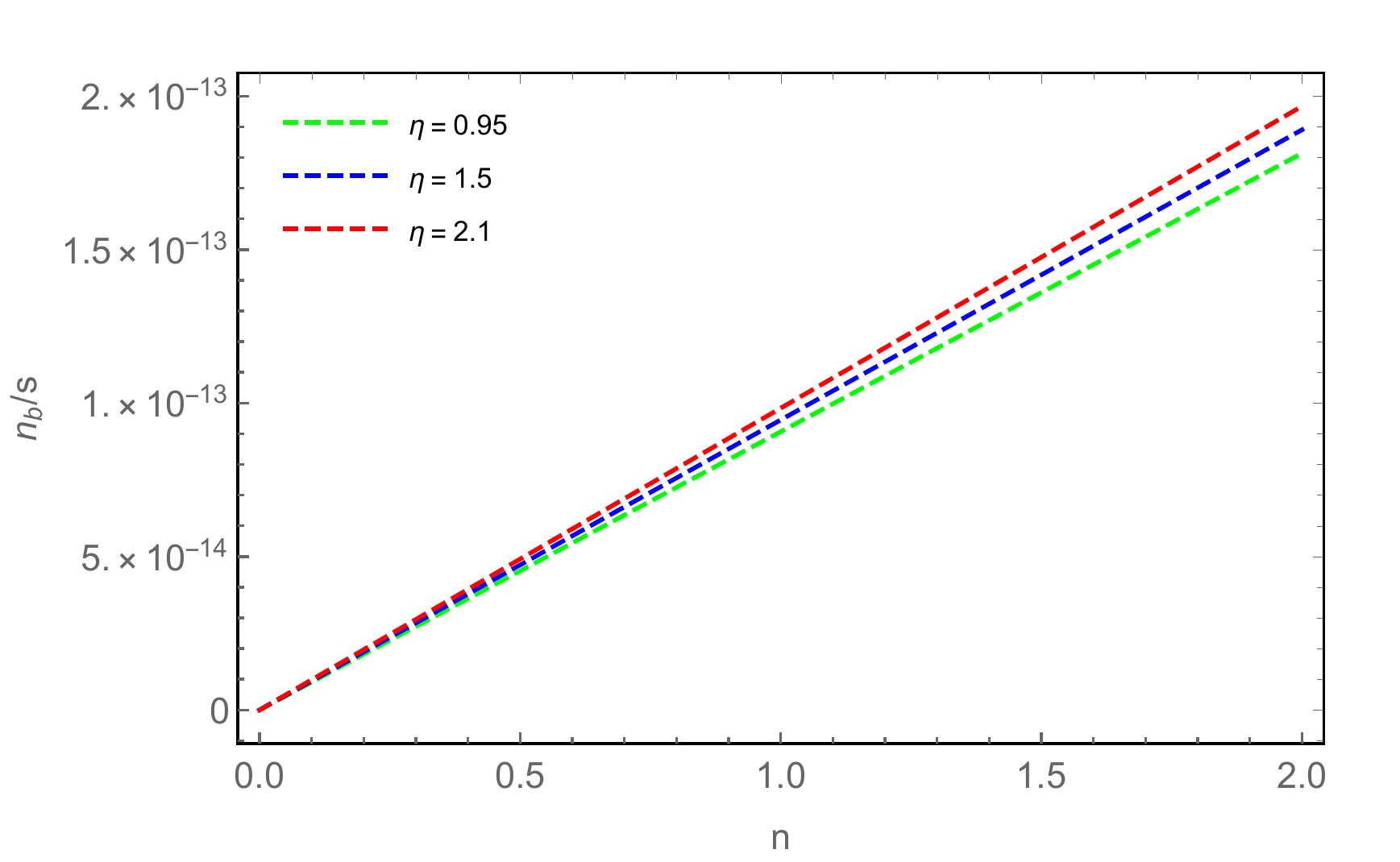}
\caption{$\frac{n_b}{s}$ vs $n$ with $\gamma=2.8$, $M_*=10^{12}$, $T_D=2 \times 10^{16}$, $\mathcal{H}_0=6.3 \times 10^{12}$}
\endminipage 
\minipage{0.50\textwidth}
\centering
\includegraphics[width=75mm]{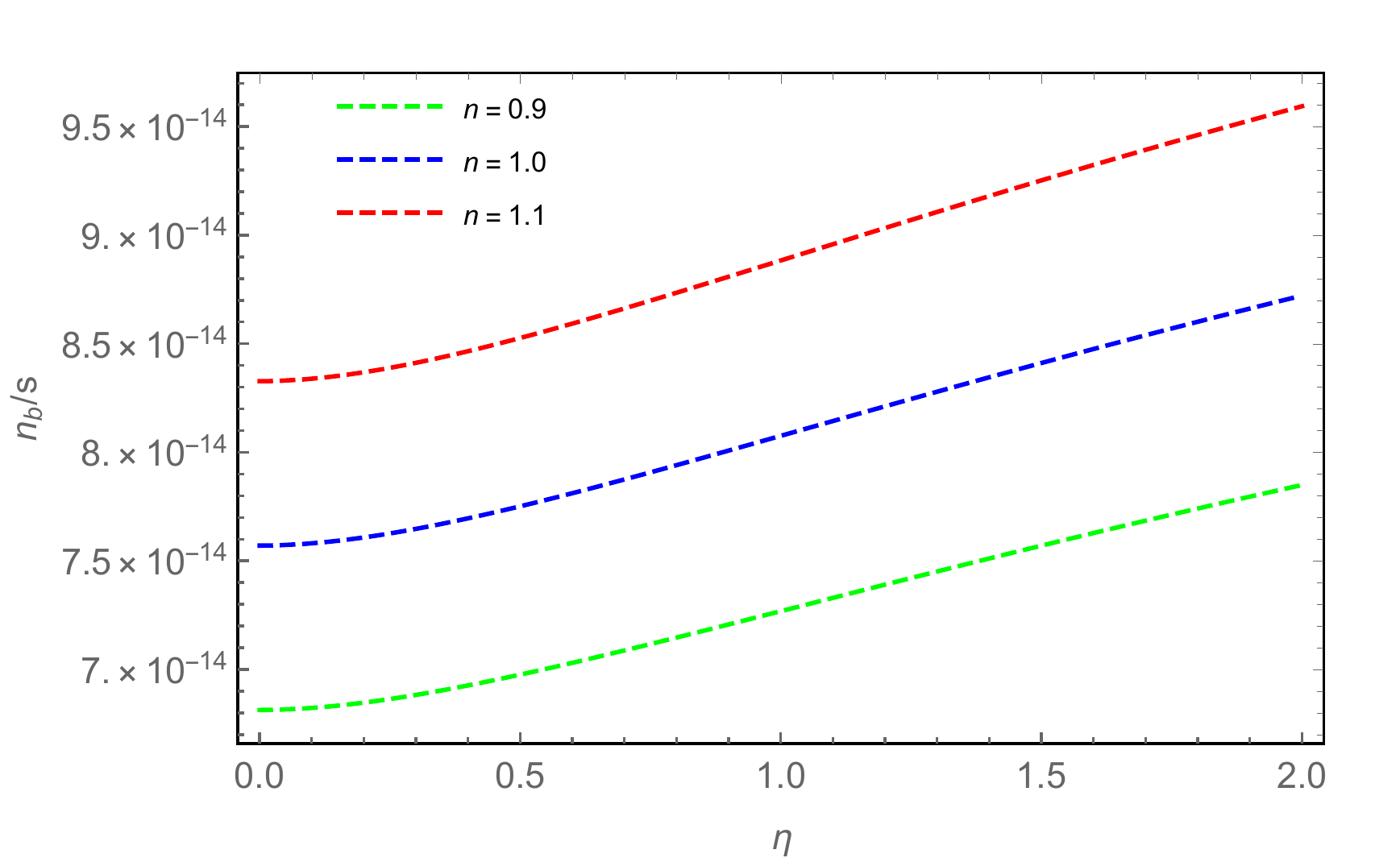}
\caption{$\frac{n_b}{s}$ vs $\eta$ with $\gamma=2.8$, $M_*=10^{12}$, $T_D=2 \times 10^{16}$, $\mathcal{H}_0=6.3 \times 10^{12}$}
\endminipage\hfill
\end{figure}

\subsection{Model II}
As a second attempt to understand the behavior of the $\frac{n_b}{s}$, in this case, we have considered the functional $f(R)$ in the form,
\begin{equation}\label{eqn.19}
f(R)=k_{1}\left(\dfrac{R}{\chi_{1}}\right)^{-\frac{\beta \mathcal{H}_{1}}{2(3+\beta)}},
\end{equation} 
where $k_1\simeq\mathcal{H}_1^\frac{\beta\mathcal{H}_1}{3+\beta}$ and $\chi_1=2\xi[(3+\beta)(\eta+2)-(\eta^2+2\eta+3)\xi]$ such that the Hubble parameter can be obtained as, $H=\frac{\mathcal{H}_1}{(3+\beta)(t_d-t_0)+1}$. It can be noted here that the corresponding deceleration parameter does not depend on time. It is controlled by the parametric value of $\beta$. Subsequently we can obtain,

\begin{eqnarray}\label{eqn.20}
\frac{n_{b}}{s}&\simeq& \left(\epsilon\frac{15g_{b}}{2\pi^{2}g_{s}}\right)\left(\frac{\chi_{1}}{M_{*}^{2}T_{D}}\right)\left(\dfrac{\chi_{1}}{2\alpha^{2}(1+2n+3n^{2})-2\alpha(1+2n)}\right)^\frac{1}{2}\left[\frac{15\beta \mathcal{H}_{1}k_{1}}{(3+\beta)g_{s}\pi^{2}T_D^{4}}\right]^{\frac{3(3+\beta)}{\beta \mathcal{H}_{1}}} \nonumber \\
&&\left(\frac{3+\beta}{\beta \mathcal{H}_{1}}+1 -\frac{(2n+1)}{2(\alpha(1+2n+3n^{2})-(1+2n))}\left({\frac{\beta \mathcal{H}_{1}}{(3+\beta)}+1+\alpha(1+2n)}\right)\right)^{\frac{3(3+\beta)}{\beta \mathcal{H}_{1}}}.
\end{eqnarray}

Incorporating the accepted values of $ \epsilon= 1, g_b$ and $g_s$, eqn. \eqref{eqn.20} can be reduced to

\begin{eqnarray}\label{eqn.21}
\frac{n_{b}}{s}&\simeq& \left(0.007168\right)\left(\frac{\chi_{1}}{M_{*}^{2}T_{D}}\right)\left(\dfrac{\chi_{1}}{2\alpha^{2}(1+2n+3n^{2})-2\alpha(1+2n)}\right)^\frac{1}{2}\left[\frac{15\beta \mathcal{H}_{1}k_{1}}{(3+\beta)g_{s}\pi^{2}T_D^{4}}\right]^{\frac{3(3+\beta)}{\beta \mathcal{H}_{1}}} \nonumber \\
&&\left(\frac{3+\beta}{\beta \mathcal{H}_{1}}+1 -\frac{(2n+1)}{2(\alpha(1+2n+3n^{2})-(1+2n))}\left({\frac{\beta \mathcal{H}_{1}}{(3+\beta)}+1+\alpha(1+2n)}\right)\right)^{\frac{3(3+\beta)}{\beta \mathcal{H}_{1}}}.
\end{eqnarray}
Fig. 3 and Fig. 4 respectively depict the behaviour of baryon number to entropy ratio with respect to $n$ and aniotropic parameter $\eta$. 

\begin{figure}[h!]
\minipage{0.50\textwidth}
\includegraphics[width=75mm]{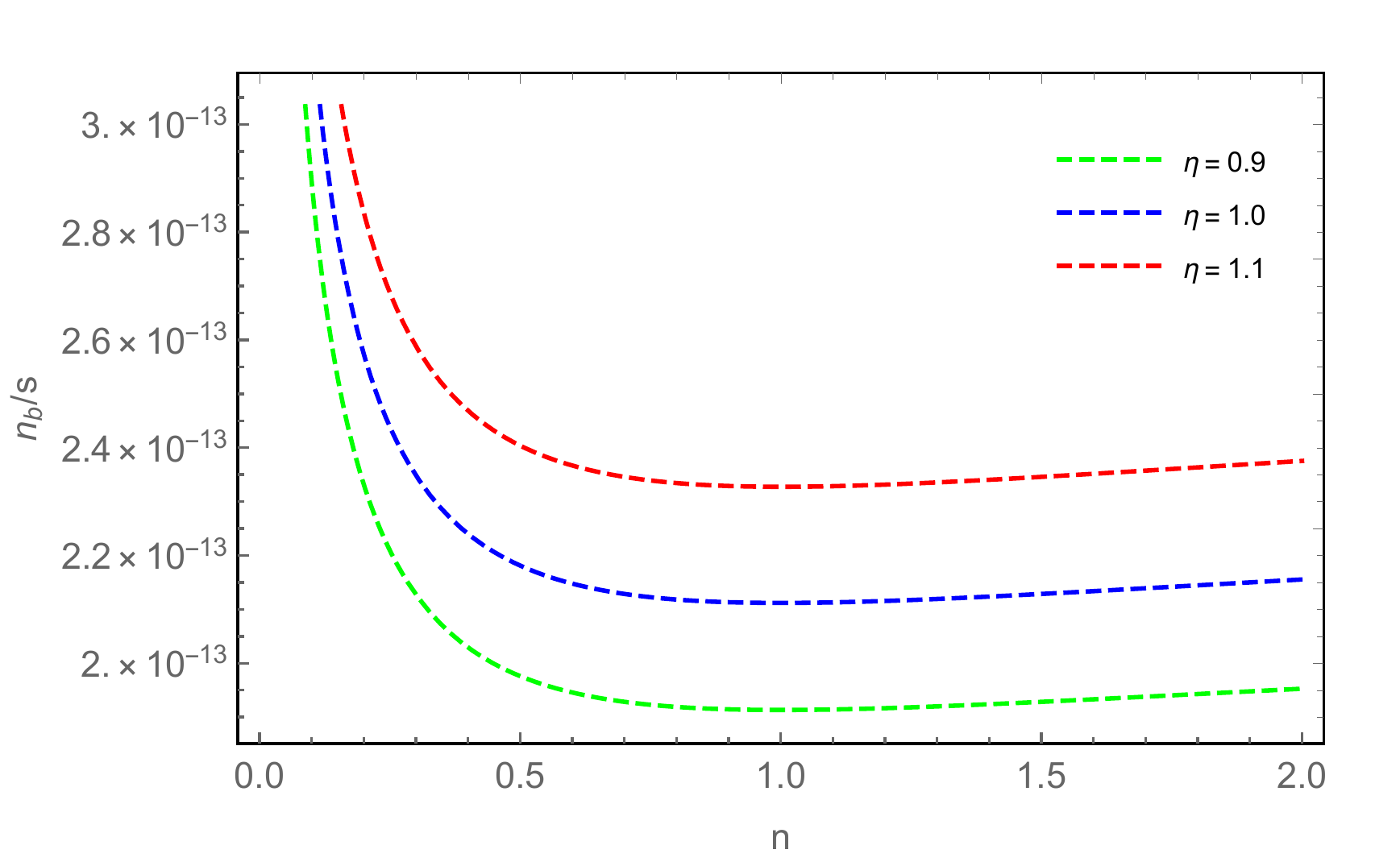}
\caption{$\frac{n_b}{s}$ vs $n$ with $\xi=1.2$, $\beta=-3.1$, $M_*=10^{12}$, $T_D=2\times 10^{16}$, $\mathcal{H}_1=6.3\times 10^{9}$}
\endminipage 
\minipage{0.50\textwidth}
\centering
\includegraphics[width=75mm]{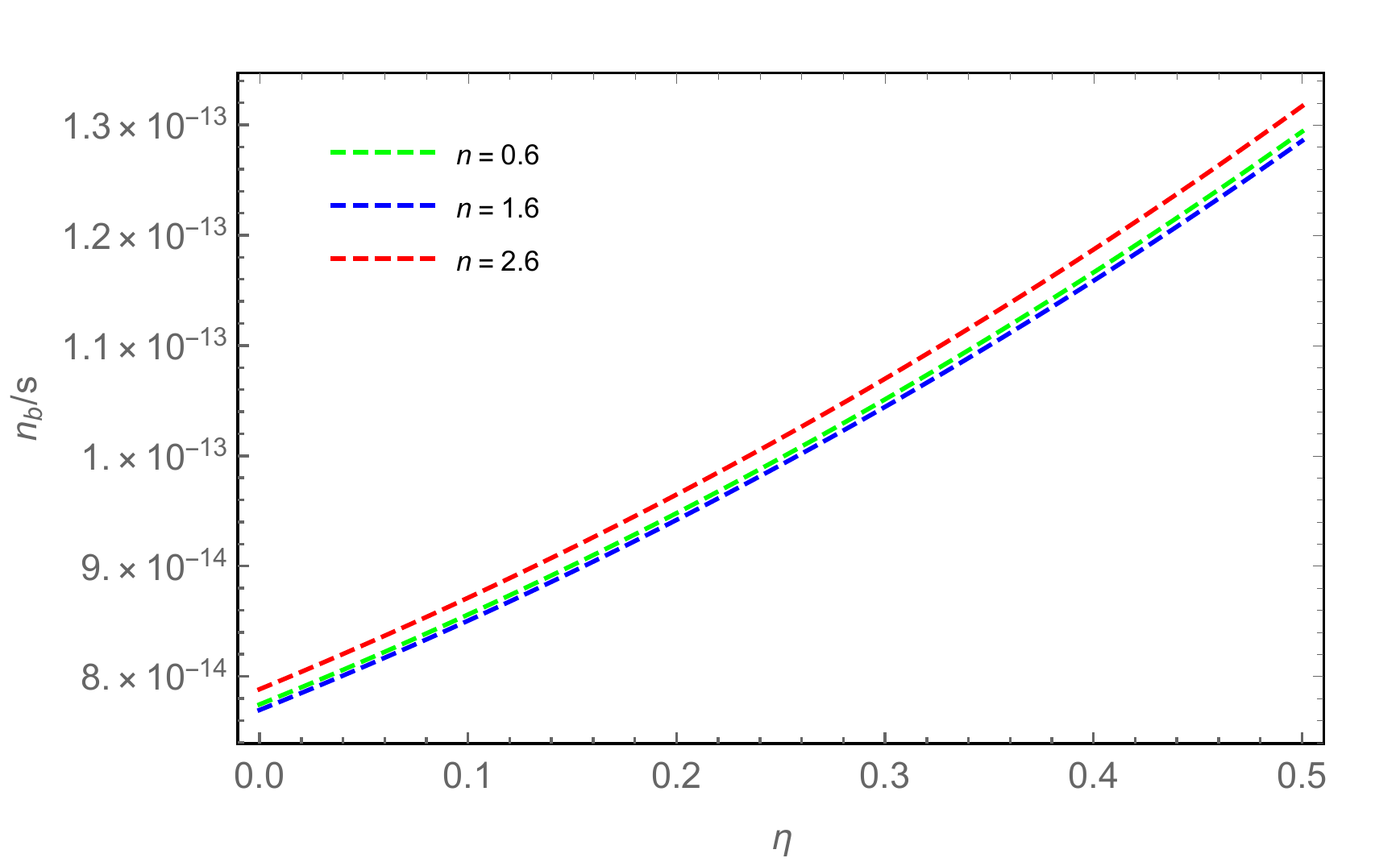}
\caption{$\frac{n_b}{s}$  vs $\eta$ with $\xi=1.2$, $\beta=-3.1$, $M_*=10^{12}$, $T_D=2\times 10^{16}$, $\mathcal{H}_1=6.3\times 10^{9}$}
\endminipage\hfill

\end{figure}
With the representative values of the anisotropic parametric value $\eta=0.9,1.0,1.1$, the ratio starts from a higher value, keep on decreasing and remains stable along a fixed value of the model parameter. Also,  for a higher value of the anisotropic parameter the ratio remains high. The $\frac{n_b}{s}$ ratio shows an increasing behaviour with respect to an increase in the anisotropic parameter for a given value of the model parameter $n$.

\section{Models Comparison and Discussions} 

The main motivation of this work lies on the compatibility of $f(R)$ theory of gravity in the study of gravitational baryogenesis in an anisotropic background of space-time. The main features of the models are,
\begin{itemize}
\item We have chosen two specific form of Ricci scalar $R$, which is the functional form of $f(R)$. The main motive to choose this forms of $R$ is that its first derivative $f_R$ of the functional depends on the mean Hubble rate, thereby gives some scope to study the anisotropic behaviour. The results mostly depend on the solution of the field equation with the assumed form.
\item The behaviour of baryon-to-entropy ratio for both the models with respect to the model parameter $n$ are almost opposite. In model-I, its starts from a null value and keeps on increasing linearly with the increase in the value of model parameter whereas in model II it starts from higher positive value and keeps on staying near $x$-axis throughout. This might be due to the behavior of $\dot{R}$ in both the models.
\item  The behaviour of baryon-to-entropy ratio with respect to the anisotropic parameter $\eta$ remains alike. The only difference in the behaviour of the baryon-to-entropy ration is the slope of the curve with respect to the anisotropic parameter. In model II, the baryon-to-entropy ratio has a larger slope with $\eta$ compared to that in model I.

\item The numerical values of baryon-to-entropy ratio in both the models are very small thereby it indicates that the universe contains a little more amount of matter than anti-matter.   
\end{itemize} 

One important aspect regarding the baryon and lepton number in presence of the electroweak fermion number violation may be discussed. There is a possibility that electroweak processes might lead to very rapid baryon and lepton number violation shortly after the electroweak phase transition ($T_c \simeq 100- 300$ GeV) which may lead to the washing out of the baryon asymmetry of the Universe \cite{Kuzmin1987, Arnold1987}. This  renders the baryogenesis mechanism as impotent. It is required that, the Universe should possess a non-zero value of B - L prior to the fermion-number violation epoch if baryon and lepton asymmetries are to survive. Also,  a non-zero value of B + L persists after the fermion-number violation epoch even though electroweak processes violate B + L. For a temperature below the electroweak phase transition, $T\leq T_c$, the lepton number may be expressed as $L_i=-\frac{41}{60}B$. At the time of primordial nucleosynthesis, the asymmetry between electron neutrinos and anti-electron neutrinos may be expressed as $L_{\nu_e}\simeq -1.6 B$. This expression is also valid for $T > T_c$. In the present work, we have not explicitly obtained the numbers associated with different quark and lepton degrees of freedom, rather, we are interested in the effect of the anisotropic parameters on the baryon-to-entropy ratio. In this context, we have skipped a detailed discussion on the baryon and lepton number conservation during the fermion number violation process.

In summary, we have presented some anisotropic models to study the gravitational baryogenesis in the framework of $f(R)$ theory of gravity by assuming some viable functional forms of $f(R)$. These functional forms have been investigated earlier to obtain viable anisotropic cosmological solutions\cite{Tripathy16}. The field equations are simplified by incorporating an anisotropic relation between the Hubble parameters along different directions. The baryon number to entropy ration number have been obtained in both the models. The effect of anisotropic parameter on the baryon-to-entropy ratio is investigated. Also, we studied the effect of the model parameter on the baryon-to-entropy ratio. We obtained a non-vanishing baryon-to-entropy ratio for both the models which increases with an increase in the cosmic anisotropy. It can be mentioned here that in comparison to the isotropic background, in an anisotropic background the baryon asymmetric is  more.  

\section*{Acknowledgement} AMA acknowledges the financial support provided by University Grants Commission (UGC) through Junior Research Fellowship (File No. 16-9 (June 2017)/2018 (NET/CSIR)),  to carry out the research work. The concept of the present collaborative work emerged from a discussion in the workshop on Relativity, Cosmology and Astrophysics held at Indira Gandhi Institute of Technology, Sarang, India during 27th-31st January 2020. The authors are thankful to the anonymous reviewer for the constructive comments and suggestions to improve the quality of the paper.

\end{document}